\documentclass[authoryear,preprint,review,12pt]{elsarticle}

 \usepackage{graphicx}
\usepackage{amssymb}

\journal{Icarus}

\begin{document}

\begin{frontmatter}



\title{Imaging polarimetry of Comet C/2012 L2 (LINEAR)}


\author[hsd]{P. Deb Roy\corref{cor1}}
\ead{pari.hkd@gmail.com}
\author[hsd]{H. S. Das\corref{cor2}}
\ead{hsdas@iucaa.ernet.in}
\author[bjm]{Biman J. Medhi}
\ead{biman@aries.res.in}

\cortext[cor1]{Corresponding author}
\cortext[cor2]{Principal corresponding author}

\address[hsd]{Department of Physics, Assam University, Silchar 788011, India}
\address[bjm]{Aryabhatta Research Institute of Observational Sciences, Manora Peak, Nainital 263129,  India}

\begin{abstract}
We present the polarimetric results and analysis of comet  C/2012 L2 (LINEAR) observed at 31$^\circ$.1
phase angle  before perihelion passage. The observations
of the comet were carried out using ARIES Imaging Polarimeter (AIMPOL)
mounted on the 1.04-m Sampurnanand telescope of ARIES, Nainital,
India on  11 and  12  March,  2013 using R photometric band ($\lambda$ =  630 nm, $\Delta$$\lambda$ =120nm).
The extended coma of the comet ($\sim65000$ km) shows a significant  variation in  the  intensity as well as polarization  profile in  all
considered directions which suggest that the dust particles originate from the active areas of the nucleus.
 The  elongation of the coma  is prominent  along  the Sun-comet  position
angle. The polarization of Comet C/2012 L2 (LINEAR) does not show steep radial dependence on the aperture size
during both the nights of observation.
A jet extended in the antisolar direction is well observed in both intensity and polarization map.
\end{abstract}


\end{frontmatter}

% \linenumbers


\section{INTRODUCTION }
\label{sec:intro}
Comet  C/2012 L2 (LINEAR) was  discovered by  the Lincoln  Near Earth
Asteroid Research  (LINEAR) Survey on  1st June, 2012. The perihelion and aphelion distance of the comet are about 1.51 au and 1166.64 au respectively  from  the Sun$^1$. The Comet C/2012 L2 (LINEAR) may belong to inner Oort-cloud  family
since the inner Oort-cloud  extends as close
as 1000 au. The comet came close to the
Earth (1.74 au) on the 26th January, 2013. The visual  magnitude of the comet  was 13.65 and  13.64 on 11
and 12 March, 2013\footnote{http://www.minorplanetcenter.net/iau/mpec/K13/K13F47.html}.

    The polarimetric study of comet gives a well defined idea about the physical properties of cometary dust grains. Numerous comets have been studied including Oort-cloud objects over a wide range of phase angle and wavelength in past through imaging polarimetry to understand features in the coma.  Recently, an Oort-cloud comet C/2009 P1 (Garradd) has been studied by \cite{dmw} and \cite{hsl} at phase angle ($\sim$ 28-35$^{\circ}$) to obtain the intensity and polarization profiles of the comet. A strong jet feature with a uniform aperture polarization has been observed by both the investigators along with a noticeable variation in intensity and polarization profile in all considered directions of the coma.

  To explain the observed photopolarimetric characteristics of comets, modeling of comet dust has been proposed by many researchers which helped a lot in understanding comet dust properties in detail (e.g., \cite{ds}, \cite{kkm}, \cite{btb}, \cite{lml}, \cite{dsk, ddp, dds, dsm, dps}, \cite{sdjb},  \cite{ll, llh}, \cite{dsa}, \cite{zmv} etc.).  Most importantly no polarimetric study of dynamically new Comet C/2012 L2 (LINEAR)
 has been reported so far. In this paper, we present the results of optical polarimetric study of this comet at 31$^{\circ}$.1 phase angle before perihelion passage.
The organization of the paper is  as follows: in Section 2 we present observation  and  data reduction,  in  Sections  3  and 4  results  and
 discussion are presented and conclusion in Section 5.

\section{OBSERVATION AND DATA REDUCTION}
The optical imaging polarimetric observation of the Comet C/2012 L2 (LINEAR) was carried out on March 11 and 12, 2013 using  1.04-m Sampurnanand Telescope of Aryabhatta Research Institute  of Observational  SciencES  (ARIES), Nainital, India (AST:  lat.=29$^\circ$22$'$N,
long.= 79$^\circ$27$'$E,  altitude=  1951m). The 1.04 m Sampurnanand  Telescope has a Cassegrain focus with a focal ratio of  f/13. The focal plane instrument used was the ARIES  Imaging Polarimeter (AIMPOL) (\cite{bim}, \cite{jep}) which consists of  a Wollaston
prism  used   to  split  the   incident  unpolarized  beam   into  two
orthogonally  polarized ordinary  and extraordinary  components  and a
rotatable half-wave plate (HWP) used to alter the polarization state of the
light wave.  The  observations of Comet C/2012 L2 (LINEAR) were carried
out in  broad band R  filter ($\lambda$ = 630  nm, $\Delta$$\lambda$
=120nm) on 11th  and 12th of March, 2013. A CCD camera of
1024 $\times$ 1024 pixels was  used during the observation with a resolution of 1.73 arcsec per pixel and the field of view is
about 8 arc minute  diameter on the sky. The gain and  the read-out noise
of the CCD are 11.98  e$^-$/ADU and 7.0 e$^-$, respectively. Four sets of exposures of each 60 s were made on both the nights of observation.The detailed description of the instrument is given in
\cite{rjp} and \cite{bim, bmb}.

\subsection{Observational procedure}
The geometrical parameters during the  observation on March
11 and March  12, 2013 are presented in Table 1. These parameters are
collected from NASA-JPL's HORIZONS system. We observed both high polarized star (HD251204) and unpolarized star (HD65583) to find out the position of polarization plane and the instrumental polarization. The observed standard stars were taken from \cite{ser}, \cite{tur} and HPOL\footnote{http://www.sal.wisc.edu/HPOL/tgts/HD251204.html}. The  data for these stars  are
depicted in  Table 2. The standard star's polarization (p), position  angle $(\theta)$ from the literature and their observed value of polarization $({  p}_{obs })$ and  position angle $({  \theta}_{obs  })$ are given  in  the fourth, fifth, sixth and seventh column in the Table 2. Since  the zero position of  HWP is not  perfectly aligned with the north-south direction  it is very much important  to determine the
offset angle properly for the  better accuracy of the result using the
relation, ${ \theta}_{o }$ = ($\theta - { \theta}_{obs }$) and the same is given in the eighth column of the Table 2.

  %%%%%%%%%%%%%%%%%%%%%%%%%%%%%%%%%%%%%% TABLE-1%%%%%%%%%%%%%%%%%%%%%%%%%%%%%%%%%%%%%%%%%%%%%%%%%

%% --Tables--
{
\renewcommand{\baselinestretch}{1}
\small\normalsize

\begin{table}
\caption{Log of the observation at AST. UT  date,  geocentric distance  ($\Delta$),
  heliocentric distance (r),  apparent visual magnitude (${\ m}_{v}$),
 phase angle ($\alpha$), extended  sun - comet radius vector position
 angle ($\phi$), projected  diameter for 1 pixel (D),  filter used in
  the observation and the exposure time of the observation (time taken
 for   1  exposure   $\times$   number  of   exposures during   the
  observation.)}
\begin{center}
\hspace*{-1cm}
\begin{tabular*}{1.2\textwidth}{@{} c c c c c c c c c@{}}
\hline \hline
  \ UT date               & \ $\Delta$        & \ r                  & \ ${\ m}_{v}$
                     &   \  $\alpha$ &   \ \ $\phi$  &  D    & Filters         &   Exposure \\

                   &   (au)        & (au)                  & ~
         & ($^\circ$)             & ($^\circ$)               & ~  (km\ pixel$^-$$^1$)                & ~
         & ~ Time \\

\hline
 March 11, 2013 & 1.92 & 1.71  & 13.65 & 31.1 & 67.8 & 2404 & R & 60 s $\times$ 4  \\
 March 12, 2013 & 1.93 & 1.70  & 13.64 & 31.1 & 68.1 & 2412 & R & 60 s $\times$ 4 \\
\hline
\end{tabular*}
\end{center}

\end{table}
}
%%%%%%%%%%%%%%%%%%%%%%%%%%%%%%%%%%%%%%%%%%%%%%%%%%%%%%%%%%%%%%%%%%%%%%%%%%%%%%%%%%%%%%%%%%%%%%%%%%%%%%%%%
%%%%%%%%%%%%%%%%%%%%%%%%%%%%%%%%%%%%%%%%% TABLE-2 %%%%%%%%%%%%%%%%%%%%%%%%%%%%%%%%%%%%%%%%%%%%%%%%%%%%%%%
{
\renewcommand{\baselinestretch}{1}
\small\normalsize

 \begin{table*}
\caption{Linear polarization of the standard polarized and unpolarized star in R-filter.
Polarization(p) and position angle ($\theta$) are from the literature, ${\ p}_{obs}$
and ${\theta}_{obs}$ for the measurements, ${\theta}_{o}$ is the offset angle.}
\begin{center}
\hspace*{-1.8cm}
\begin{tabular*}{1.25\textwidth}{c c c c c c c c c}
\hline \hline
 \ UT date   & \ Filters   & \ Standard Star   & \ p(\%)        & \ ($\theta$)
        & \  ${\ p}_{obs}$(\%) & \  ${\theta}_{obs}$ &   \ ${ \theta}_{o }$   \\
\hline

 March 11, 2013 & R & HD251204 & 4.79$\pm$0.4  & 152.9 & 4.85 $\pm$ 0.30 & 154.9 $\pm$ 1.7 & -2 \\

 March 11, 2013 & R & HD65583 & 0.05  & 149.3 & 0.12 $\pm$ 0.18  & 146 $\pm$ 41  & 3.3   \\
\hline

 March 12, 2013 & R & HD251204 & 4.79$\pm$0.4  & 152.9 & 4.86 $\pm$ 0.32  & 154.9  $\pm$ 1.7  & -2 \\
 March 12, 2013 & R & HD65583 & 0.05  & 149.3 & 0.13 $\pm$ 0.18 & 146 $\pm$ 40 & 3.3   \\
\hline
\end{tabular*}\label{tab2}
\end{center}

\end{table*}
}
%%%%%%%%%%%%%%%%%%%%%%%%%%%%%%%%%%%%%%%%%%%%%%%%%%%%%%%%%%%%%%%%%%%%%%%%%%%%%%%%%%%%%%%%%%%%%%%%%%%%%%%%%%%%
    The observed  images are analyzed using standard IRAF routines. Each image is  bias and flat field corrected. The photometric  center of  each observed images  has been  found out  with  a precision  of  0.1  pixel. The Wollaston prism and the rotatable half-wave plate allow to form the two orthogonally polarized components of a single object in the CCD camera. The set of images for each orientation of the fast axis are combined after the proper alignment to increase the  signal to noise ratio of each polarized component. The data reduction procedures are systematically discussed in \cite{dmw}.

   \section{RESULTS}

   \subsection{Intensity Images}

     The  intensity images are  produced by adding  two polarized
    components with  a proper alignment. The total  intensity is given
    by $ I= I_{e}+ I_{o}$, where $I_{e}$ is the intensity of extraordinary and $I_{o}$ is the intensity of ordinary image. The  intensity  image with  contours  along  with the  rotational gradient treated image are shown  in the Figure 1 (a--d). The position angle of
     the  Sun$-$comet  radius  vector  is  67.8$^\circ$ and  68.1$^\circ$  on
     11th and 12th March respectively.

      \  \ The radial profile through the intensity images is a diagnostic tool to study the cometary dust grains.
    To  compare the variation  in the  average radial intensity  profile
   between the two dates of  observation, the intensity is plotted against
   the photocentric distance after the images being normalized in  e$^-$/s  for  both nights  of
   observations and is  shown in Figure 2. For both the nights of observation, the  decrease in intensity is well noticed
 when the photocentric distance is increased radially. The profiles are dominated by the seeing when
 the photocentric distance is less than 4 arcsec ($\sim$5500 km).

    The intensity along the different directions explores the various evolutionary mechanisms working in the cometary coma. It  is
    noticed that the intensity is  higher in the tailward direction as
    compared  to the solar  direction. We  also analyzed  the intensity
    profile  in  north-west  and   south-east  direction  in  which  a
   variation in the profile is noticed. The variation of the  profile can be well inferred from
   the change  in the  slope value delimitated  in Table-3. The slope varies from $-0.85$ to $-1.61$ and has also variations between different directions in the coma.

     The change in the slopes is a regular phenomenon for an isotropic coma. The variation in the intensity profile is mainly due to the solar radiation pressure which sorts the particles according to their cross-section and mass. The sublimation of ice or organic material coated grains may change the physical properties of the dust particles as they recedes from the nucleus which would also lead to this variation.

      \ \ The elongation of  the extended  coma is  prominent along the  Sun--comet radius vector. The intensity  image  is  treated  with  the
   Larson-Sekanina's rotational gradient  technique (\cite{lar}) to find
   the special feature  like jet activity present  in the comet. It can be noticed from Fig. 1c and d that a jet is extended in
      the antisolar direction on both the nights. A change in the direction of the jet between two nights is being noticed which is most likely due to the rotation of the nucleus.

%%%%%%%%%%%%%%%%%%%%%%%%%%% FIG-1 %%%%%%%%%%%%%%%%%%%%%%%%%%%%%%%%%
\begin{figure*}
%\vspace{2cm}
%\hspace{-0.4cm}
\begin{center}
\includegraphics[width=90mm]{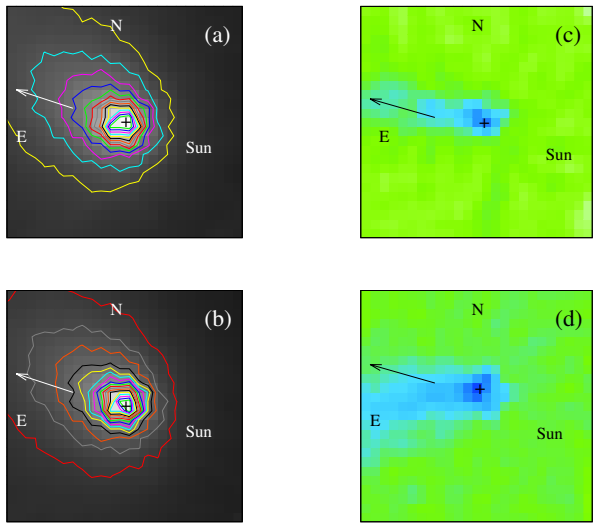}
  %% to include a figure, or

  %% to leave a blank space
  \caption{Intensity map  with contours for
  (a)  11th  March and (b)  12th  March,  2013.  The false color rotational  gradient
  treated image  for (c) 11th March and (d) 12th March,  2013. The '$+$'
  mark denotes the photocenter of the comet. The arrow shows the position angle of the Sun$-$Comet radius vector which is 67$^\circ$.8 and 68$^\circ$.1 on 11th and
  12th March respectively. Scale: 60000km $\times$ 60000km.}
\end{center}

\end{figure*}

%%%%%%%%%%%%%%%%%%%%%%%%%%%%%%%%%%%%%%%%%%%%%%%%%%%%%%%%%%%%%%%%%%%%%
%%%%%%%%%%%%%%%%%%%%%%%%% FIG-2 %%%%%%%%%%%%%%%%%%%%%%%%%%%%%%%%%%%%
\begin{figure}
%\vspace{1.9cm}
%\hspace{0cm}
\begin{center}
\includegraphics[width=85mm]{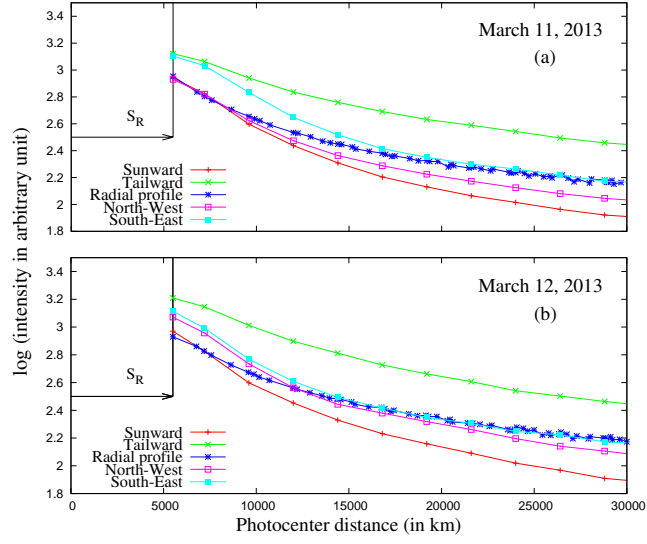}
  %% to include a figure, or

  %% to leave a blank space
\caption{Cuts through the coma sunward, tailward, south-east, north-west
and radial profile for two nights of observation. Vertical line represents
the seeing radius (${S}_{R}$) limit for each observed night.}
\end{center}
\end{figure}
%%%%%%%%%%%%%%%%%%%%%%%%%%%%%%%%%%%%%%%%%%%%%%%%%%%%%%%%%%%%%%%%%%%%%

%%%%%%%%%%%%%%%%%%%%%%%%%%%%%%%%%%%%%%% TABLE-3 %%%%%%%%%%%%%%%%%%%%%%%%%%%%%%%%%%%%%%

{
\renewcommand{\baselinestretch}{1}
\small\normalsize
\begin{table*}
\caption{Intensity profile variation throughout the coma.}.
\begin{center}
\hspace*{-2cm}
\begin{tabular*}{1.2\textwidth}{c c c c c c c c c}
\hline \hline
Radial distance to the          & \    & \ 9000        & \ 12000
         &   \  15000 &   \ 18000  & \ 21000 &   \ 24000  & \ 30000      \\

     photometric center (in km)      &        &           & ~                  & ~
         & ~             & ~         & ~             & ~                  \\

\hline
\underline{\textbf{Solar direction}}\\
 March 11, 2013 &  & -1.40  & -1.47 & -1.50 & -1.50
  & -1.51 & -1.52 & -1.53 \\
 March 12, 2013 &  & -1.50 & -1.57 & -1.58 & -1.60
  & -1.61 & -1.61 & -1.62   \\

  \hline
\underline{\textbf{Tailward direction}}\\
 March 11, 2013 &  & -0.85  & -0.92 & -1.00 & -1.10
  & -1.14  & -1.16 & -1.20\\
 March 12, 2013 &  & -0.90 & -1.05 & -1.15 & -1.20
  & -1.25 & -1.30 & -1.31   \\

  \hline
\underline{\textbf{North-West direction}}\\
 March 11, 2013 &  & -1.30  & -1.43 & -1.41 & -1.26
  & -1.20 & -1.16 & -1.10  \\
 March 12, 2013 &  & -1.45 & -1.54 & -1.58 & -1.29
  & -1.23 & -1.15 & -1.00   \\

  \hline
\underline{\textbf{South-East direction}}\\
 March 11, 2013 &  & -1.20  & -1.48 & -1.52 & -1.54
  & -1.30 & -1.18 & -1.10  \\
 March 12, 2013 &  & -1.47 & -1.51 & -1.54 & -1.54
  & -1.40 & -1.31 & -1.05   \\
  \hline

\end{tabular*}\label{tab2}
\end{center}
\end{table*}
}
%%%%%%%%%%%%%%%%%%%%%%%%%%%%%%%%%%%%%%%%%%%%%%%%%%%%%%%%%%%%%%%%%%%%%%%%%%%%%%%%%%%%%%%%%%%%%%

\subsection{Linear polarization}

%%%%%%%%%%%%%%%%%%%%%%%%%%%%%%%%%%%%%%% TABLE-4 %%%%%%%%%%%%%%%%%%%%%%%%%%%%%%%%%%%%%%%%%
{
\renewcommand{\baselinestretch}{1}
\small\normalsize
\begin{table*}
\caption{Linear polarization in percent at different apertures . UT date, Filters used in the observation, aperture diameter (D) and the observed position angle of the polarization vector (${ \theta}_{obs }$).}
\begin{center}
\hspace*{-2.35cm}
\begin{tabular*}{1.38\textwidth}{c c c c c c c c c}
\hline \hline
D (in km)         & \     & \ 12000   & \ 20000   & \ 30000        & \ 40000
         &   \  50000 &   \ 60000  & \ ${ \theta}_{obs }$ \\

     UT date   & Filter \\

\hline
 March 11, 2013 & R & 2.3$\pm$0.4 & 2.6$\pm$0.3 & 2.6$\pm$0.3  & 2.6$\pm$0.3 & 2.6$\pm$0.3 & 2.8$\pm$0.3
  & 160.3$\pm$3.3  \\
 March 12, 2013 & R & 2.4$\pm$0.3 & 2.4$\pm$0.3 & 2.2$\pm$0.3  & 1.9$\pm$0.3  & 2.0$\pm$0.3 & 1.9$\pm$0.2
  & 148.1$\pm$4.4 \\
  \hline

\end{tabular*}\label{tab2}
\end{center}

\end{table*}

}

%%%%%%%%%%%%%%%%%%%%%%%%%%%%%%%%%%%%%%%%%%%%%%%%%%%%%%%%%%%%%%%%%%%%%%%%%%%%%%%%%%%%%%%%%%%%%%%%%%

\subsubsection{Aperture polarization}

The aperture  polarization values are  estimated from
the  integrated  flux  measured  corresponding to  all  the  polarized
components   through  increasing   apertures   from  the   photocenter.

Since the comet is not so bright, four different exposures of 60 s at R-filter are made on both nights of
observation to increase the signal to noise ratio.  Then  all  the  images  corresponding  to  a  particular
angle of rotation   of    HWP   are   combined   to    build   a   polarization
component. Thus four polarized components are produced for four different angles of rotation $0^{\circ}$, $22.5^{\circ}$,  $45^{\circ}$ and $67.5^{\circ}$ of HWP. Finally with all the properly  aligned polarization components, the aperture linear  polarization values are estimated.  The polarization  (${\ p}_{obs}$)   corresponding  to  different   apertures  and  the
polarization  angles  (${  \theta}_{obs  }$)  obtained  for  both nights of observation are shown in Table 4. The polarization values obtained
for  comet  C/2012 L2 (LINEAR) is  found to be almost uniform with  the change of aperture
which is pointing towards significant dust domination that overpowers the influence of gas.

It has been noticed that a change in the aperture polarization between the two dates is being observed with values typically ranging from ($2.8 \pm 0.3$) per cent to ($1.9 \pm 0.2)$ per cent in the outer coma. This change may happen due to the variation in the jet activity, the possible gaseous contamination through the broadband red filter ($\Delta \lambda$ = 120 nm) which results in depolarization and also due to
the change in dust properties of the comet.

\subsubsection{Polarization map}

Polarization maps  have been constructed with the four  properly aligned
polarized components for March 11 and 12, 2013 at R filter and is shown in  Figure 3.
  The higher polarization of about 5\% is
  noticed  in the  near  nucleus region. The  polarization is  found to  be uniform
within  the field  of  view of  about  10,000 km  near the  photocenter on  both nights  of
observation which shows the uniformity in the dust properties in the near nucleus region. The variation of
polarization between 2\% and 1\%  is observed in the outer coma of the comet. A strong jet is also noticed in the polarization map with a slight extension in the antisolar direction which shows higher polarization of 3--4\% as compare to the surrounding coma on  both the nights.
%%%%%%%%%%%%%%%%%%%%%%%%%%%%%%%%%%%%%%%%%%%%%%%%%% FIG-3%%%%%%%%%%%%%%%%%%%%%%%%%%%%%%%%%%%%%%%%%%%%
\begin{figure}[h]
\begin{center}
%\vspace{-3.7cm}
\hspace{2cm}
\includegraphics[width=110mm]{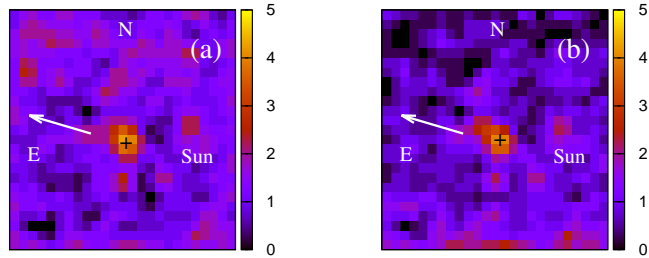}
  %% to include a figure, or

  %% to leave a blank space
\caption{Polarization maps (the levels are in \%) at 31.1$^\circ$ phase angle at R- filter on (a) March 11, 2013
and (b) March 12, 2013. The `$+$' mark denotes the photocenter of the comet. The white arrow shows the position angle of the Sun $-$ Comet radius vector which is 67$^\circ$.8 and 68$^\circ$.1 on 11th and
  12th March respectively. Scale: 60000km $\times$ 60000km. }
\end{center}
\end{figure}
%%%%%%%%%%%%%%%%%%%%%%%%%%%%%%%%%%%%%%%%%%%%%%%%%%%%%%%%%%%%%%%%%%%%%%%%%%%%%%%%%%%%%%%%%%%%%%%%%%%%%%%%%%

%%%%%%%%%%%%%%%%%%%%%%%%%%%%%%%%%%%%%%%%%%% FIG-4 %%%%%%%%%%%%%%%%%%%%%%%%%%%%%%%%%%%%%%%%%%%%%%%%%%%%%%%
\begin{figure}[h]
\begin{center}
\vspace{1.0cm}
%\hspace{0.5cm}
\includegraphics[width=75mm]{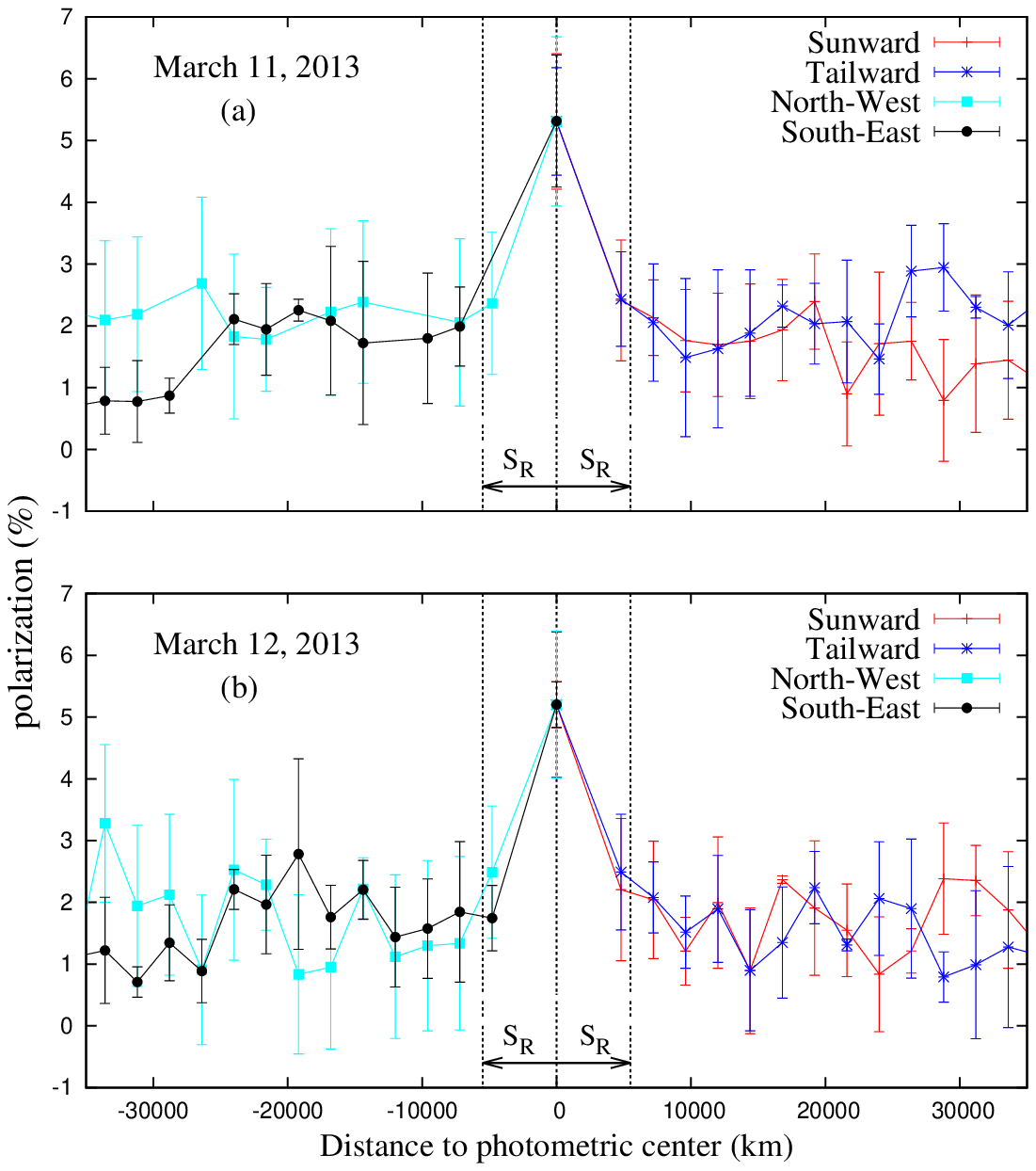}
 %% to include a figure, or

 %% to leave a blank space
 \vspace{0.5cm}
\caption{Variation of polarization profile with photocentric distance for two observed nights.
Vertical line represents the seeing radius (${S}_{R}$) limit for each observed night.  }
\end{center}
\end{figure}
%%%%%%%%%%%%%%%%%%%%%%%%%%%%%%%%%%%%%%%%%%%%%%%%%%%%%%%%%%%%%%%%%%%%%%%%%%%%%%%%%%%%%%%%%%%%%%%%%%%%%%%%%%%%%

\subsubsection{Polarization profile}

The variation in the  polarization profile with the photocentric
distance for  both the nights of  observation is shown  in Figure  4. A
significant variation  in the polarization profile is  being observed in
all  possible  directions for  both the nights. The  higher
polarization  is  noticed  in  the   inner  coma  of  the  comet. Polarization
decreases  gradually  up to a ceratain distance with  an  increase  in  the
photocentric distance  (i)  up  to 12,000 km  in  the  sunward
direction and  up to 15,000 km  in the tailward direction  on 11 March 2013
 and (ii) up to 11,000  km in the sunward and up to 14,000
km in the tailward direction is  being observed on 12 March 2013. The  polarization value in the  north-west and south-east
direction  is falling  with  the increase  in  the photocentric
distance (i)  up to  8000 km  and 12,000 km  respectively in  the first
night and  (ii) 8000  km and  13,000 km  respectively  in the
second night of  observation. Polarization is  found to vary
in  the  outer  coma of  Comet C/2012 L2 (LINEAR) in  all directions for  both
the nights of observation. The variations are within the errors when the photocentric distance is within 30,000 km. But when this distance is greater than 30,000 km, the
polarization value is being found to be  higher  in  the  tailward  direction as  compared  to
the  sunward direction on the first night of observation whereas a completely opposite trend is
being detected on the other date. The error in polarization is almost consistently lower in the
inner coma of the comet in all possible directions as compared to the slightly higher
value in the outer coma of the comet. The variation in the
polarization profile indicates the non uniformity in
the polarization  distribution which is due to change in the dust intrinsic
properties endures in different apertures in all  possible  directions
specially in the outer coma of the comet.

\section{Discussion}
The  intensity profile  of the Comet C/2012 L2 (LINEAR) yields some
 important  results  which  show   a  significant  variation  in  the
 intensity feature  of  the  comet  in all  possible  directions. The
 intensity  varies  slowly in the inner and outer coma of the Comet C/2012 L2 (LINEAR) in tailward
 direction as compare to  the sunward  direction where  the intensity
 falls steeply  in   the outer coma on both the nights of  observation. The variation in intensity is
 also   being   detected  between   the   north-west  and   south-east
 direction. The deviation of the profile in all possible directions from the standard canonical nature is due to the ongoing different evolutionary processes which collectively effect the light scattering properties of the dust grains. Temporal changes in the dust production is the primary reason for intensity variation. Sublimation of the ice or organic coated grains when they accelerated away from the nucleus due to the solar radiation pressure result in the shrinking of the grains which would also cause this variation (\cite{tlk}, \cite{fer}).
  The variation  of slopes in all direction is a feature of
 the asymmetric coma strictly directed away from the Sun.  The Comet C/2012 L2 (LINEAR) also shows a variation in the aperture polarization between the two dates with values typically ranging from ($2.8\pm0.3$) per cent to ($1.9\pm0.2)$ per cent in the outer coma. This change is due to the variation in the jet activity, the possible gaseous contamination through the broadband red filter and also due to the change in dust properties of the comet (grain size, porosity, refractive indices).  In the jet-like structures, the  polarization is generally high while it is comparatively lower in the circumnucleus halo than in the surroundings (\cite{rld}, \cite{tcd}, \cite{hl}, \cite{jg}).

Many investigators studied variation of polarization with change of aperture size for some dusty and gaseous comets at lower phase angle. \cite{rkv} studied the gaseous comet C/2001 A2 (LINEAR) which shows high positive polarization in the near nucleus region at 26$^\circ$.5 and 36$^\circ$.2 phase angle where dust concentration is very high but in the outer coma gas dominates which result in the steep fall in polarization value. The similar trend of strong radial dependence of polarization with increasing aperture size is also being observed for gaseous Comets C/1996 Q1 (Tabur) (\cite{kjr}) and 2P/Encke (\cite{kjb}, \cite{jew}). It is also observed that dusty comets do not show steep radial dependence of polarization with increasing aperture size (see, e.g., \cite{mb}). \cite{kkk} summarized published polarization data for different gas and dust rich comets and showed that they have a major difference in the polarization behavior with increasing the aperture size. Comet C/2012 L2 (LINEAR) does not show steep radial dependence of polarization on the aperture size at phase angle 31$^{\circ}$. This is also observed in dusty comet C/2009 P1 (Garradd) at comparable phase angle (\cite{dmw} and \cite{hsl}. Since no polarimetric observation has been reported so far for Comet C/2012 L2 (LINEAR) at larger phase angle, it is not possible to comment about the class of this comet at phase angle 31$^\circ$.

    Jet activity is a special feature of active comets which is prominently  detected in both the rotational gradient treated image and in the polarization map of the Comet C/2012 L2 (LINEAR) with a variation in the direction of the extension between the two nights of observation. Jet shows  high positive polarization values as compare to the whole coma polarization  on  both the nights. The high polarization usually found in the jets coming out from the nucleus is mainly due to the presence of small Rayleigh grains which are accelerated to higher speed from the surface of the nucleus by the gas drag due to their small cross-sectional area. The average polarization between 2.3\% and 2.8\% is being estimated throughout the apertures on March 11, 2013 whereas the insignificant variation in the aperture polarization between 2.4\% and 1.9\% is also being found in  the inner and the outermost coma of the Comet C/2012 L2 (LINEAR) on  March 12, 2013. Since the polarization is  sensitive to the physical properties of the cometary grains so the nonuniformity in  the polarization distribution in  all  possible directions of the cometary coma explores the different grain population.

\section{Conclusions}
\begin{enumerate}
  \item The integrated aperture polarization of the Comet C/2012 L2 (LINEAR) at phase angle 31$^{\circ}$  does not show steep radial dependence on the aperture size during both the nights of observation. Since no polarimetric observation has been reported for Comet C/2012 L2 (LINEAR) at higher phase angles, it is very difficult to comment about the class of the comet at this phase angle.

  \item The variation in the intensity profile is observed in all considered directions of Comet C/2012 L2 (LINEAR) with a change in the slope between $-0.85$ and $-1.61$ throughout the coma. The variation in the intensity profile is mainly due to the solar radiation pressure which sorts the particles according to their cross-section and mass. Further, sublimation of the ice or organic coated grains may change the physical properties of the dust grains.

  \item A prominent jet with a change in the direction of extension between the two nights is being detected in both the rotational gradient treated image and in the polarization map. The variation is most likely due to the rotation of the nucleus.
\end{enumerate}

\section{Acknowledgements}
 Through  this acknowledgment,  we  express our  sincere gratitude  to
 ARIES, Nainital for allocation of observation time. We are highly grateful to anonymous
 reviewers of the paper for their constructive comments which definitely helped to improve the quality of the paper. This work is supported by
 the Department  of Science \& Technology (DST),  Government of India,
 under SERC-Fast  Track scheme (Dy. No. SERB/F/1750/2012-13).



%% References


\bibliographystyle{model1a-num-names}




%% \harvarditem[]{}{20**} must have the following form:
%%   \harvarditem[]{}{20**}{key}...
%%

% \harvarditem[]{}{20**}{}

% \end{thebibliography}

\end{document}